\documentclass[print]{aa}

\usepackage{txfonts}
\usepackage{graphicx}
\begin{document}
\def\LSUN{\rm L_{\odot}}

\def\MSUN{\rm M_{\odot}}

\def\RSUN{\rm R_{\odot}} 

\def\MSUNYR{\rm M_{\odot}\,yr^{-1}}

\def\MSUNS{\rm M_{\odot}\,s^{-1}}

\def\MDOT{\dot{M}}

   \title{Accretion of Low Angular Momentum Material onto Black Holes: Radiation
Properties of Axisymmetric MHD Flows.}

   \author{Monika Moscibrodzka
          \inst{1,}\inst{2}, Daniel Proga \inst{2}, Bozena Czerny  
     \inst{1}, Aneta Siemiginowska \inst{3}
          }

   \institute{1. N. Copernicus Astronomical Center, 
              Bartycka 18, 00-716 ,Warsaw ,Poland\\
	      2. Department of Physics, University of Nevada, Las Vegas, NV 89154, USA\\
	      3.Harvard-Smithsonian Center for Astrophysics, 60 Garden Street, Cambridge, MA 02138, USA
                          }
\offprints{Monika Moscibrodzka \email{ mmosc@camk.edu.pl}}
   \date{Received  / Accepted  28/06/2007 }

   \abstract{

{\it Context}:
Numerical simulations of MHD accretion flows in the vicinity
of a supermasssive black hole provide important insights to the problem
of why and how systems -- such as the Galactic Center -- are underluminous
and
variable. In particular, the simulations indicate that low angular
momentum accretion flow is strongly variable both quantitatively and
qualitatively. This variability and a relatively low mass accretion rate
are caused by an interplay between a rotationally supported torus, its outflow
and a nearly non-rotating inflow.

{\it Aims}:
To access applicability of such flows to real objects, we examine
the dynamical MHD studies with computations of the time dependent radiation
spectra predicted by the simulations.

{\it Methods}:
We calculate synthetic broadband spectra of accretion flows using Monte
Carlo techniques. Our method computes the plasma electron temperature
allowing
for the pressure work, ion-electron coupling, radiative cooling, and
advection. The radiation spectra are calculated by taking into account
thermal synchrotron and bremsstrahlung radiation, self absorption,
and Comptonization processes. We also explore effects of non-thermal
electrons. We apply this method to calculate spectra
predicted by the time-dependent model of an axisymmetic MHD
flow accreting onto a black hole presented by Proga and Begelman.

{\it Results}:
Our calculations show that variability in an accretion flow
is not always reflected in the corresponding spectra, at least
not in all wavelengths. We find no one-to-one correspondence
between the accretion state and the predicted spectrum. For example, we
find that two states with different properties -- such as the geometry
and accretion rate -- could have relatively similar spectra.
However, we also find two very different states with very different spectra.
Existence of nonthermal radiation may be necessary to 
explain X-rays flaring because thermal bremsstrahlung, that dominates 
X-ray emission, is produced at relatively large radii where
the flow changes are small and slow.

   \keywords{Magnetohydrodynamics (MHD), Radiation mechanisms:general, Radiative transfer}
   }

   \maketitle


\section{Introduction}

The overall radiative output from the Galactic center (GC) is very
low for a system hosting a supermassive black hole (SMBH).
This under-performance of the nearest SMBH challenges
our understanding of mass accretion processes.

Certain properties of GC are relatively well established. For example,
the existence of a $\sim 3.7 \times 10^{6}~\MSUN$ SMBH in GC is supported by
the stellar dynamics (e.g. Schoedel et al. 2002, Ghez et
al. 2003). The nonluminous matter within 0.015 pc of GC is associated
with Sgr~A$^\ast$, a bright compact radio source~(Balick \& Brown
1974).  Observations of Sgr~A$^\ast$ in X-ray and radio bands reveal a
luminosity substantially below the Eddington limit, $L_{\rm Edd}=3
\times 10^{44}$ erg s${}^{-1}$.  In particular, {\it Chandra}
observations show a luminosity in 2-10 keV X-rays of $\approx 2\times
10^{33}$ erg s${}^{-1}$, ten orders of magnitude below $L_{\rm Edd}$
(Baganoff et al. 2003).  {\it Chandra} observations also revealed an
X-ray flare rapidly rising to a level about 45 times as large, lasting
for only $\sim 10^4$~s, indicating that the flare must originate near
the black hole~ (Baganoff et al. 2001; Baganoff et al. 2003).  A
strong variability of Sgr~A$^\ast$ was also observed in radio and
near-infrared (e.g., Eckart et al. 2005).

Despite a relative wealth of the observational data and significant
theoretical developments, a character of the accretion flow in GC is
poorly understood. In particular, it is unclear whether the emission
comes from inflowing or outflowing material (e.g., Markoff et
al. 2001; Yuan, Markoff, Falcke 2002). Although the accretion fuel is
most likely captured from the stellar winds, 
mass accretion rate $\MDOT_a$ onto the central black hole is uncertain. For example, the
quiescent X-ray emission measured with {\it Chandra} implies $\MDOT_a
\sim$ a few $\times 10^{-6}~\MSUNYR$ at $\sim 10^5 R_g$ (assuming
spherically symmetric Bondi accretion Baganoff et al. 2003) whereas
the measurements of the Faraday rotation in the millimeter band imply
$\MDOT_a~\sim$ a few $\times 10^{-7}-10^{-8}~\MSUNYR$ (Bower et
al. 2003, 2005). Recent measurements of Faraday rotation 
of Marrone et al. (2006) imply even
lower mass accretion rate $\sim 10^{-7}-10^{-9}~\MSUNYR$, where the
lower limit is valid for sub-equipartition disordered or toroidal
magnetic field.

 An increasing amount of the available
observational data of the Galactic Center has motivated a recent
development in the theory and simulations of accretion flows. 
Most theoretical work has been focused on steady state models, 
which assume that the specific angular momentum, $l$, is high and 
the inflow proceeds through a form of the so-called radiatively
inefficient accretion flow (RIAF; e.g., Narayan et al. 1998; 
Blandford \& Begelman 1999; Stone, Pringle \& Begelman 1999;
Quataert \& Gruzinov 2000; Stone \& Pringle 2001).  
A high specific angular
momentum considered in these studies is a reasonable assumption but
not necessarily required by the data. In fact, 
many observational aspects of Sgr~A$^\ast$  
can be better explained by low-$l$ accretion flows.
For example, Moscibrodzka (2006, hereafter M06) showed
that models of purely spherical accretion flows underpredict the total
observed luminosity if the flow radiative efficiency is not
arbitrarily adopted but calculated self-consistently from the
synchrotron, bremsstrahlung, and Compton emission of the accreting
material. M06's results imply that low-$l$ flows can also
reproduce the observed total luminosity as good as the high-$l$
flows.
Additionally, the intrinsic time variability of low-$l$ magnetized flows 
found by Proga \& Begelman (2003, PB03 hereafter)
can naturally account for some of Sgr~A$^\ast$ variability
(Proga 2005).

In this paper, we present synthetic continuum spectra and Faraday rotation 
measurements (RM)
calculated based on time-dependent two-dimensional (2-D)
axisymmetric MHD simulations of accretion flows performed PB03.  
PB03's simulations are for slowly
rotating flows and are complementary to other simulations which considered
strongly rotating accretion flows set to be initially
pressure-rotation supported torii (e.g., Stone \& Pringle 2001; Hawley
\& Balbus 2002 (hereafter HB02); Krolik \& Hawley 2002; Igumenshchev \& Narayan
2002). PB03 attempt to mimic the outer boundary conditions of classic
Bondi accretion flows modified by the introduction of a small,
latitude-dependent angular momentum at the outer boundary, a
pseudo-Newtonian gravitational potential, and weak poloidal magnetic
fields.
Such outer boundary conditions allow for the density distribution at
infinity to approach spherical symmetry.  Recent X-ray images taken by
the {\it Chandra} show that the gas
distribution in the vicinity of a SMBH at the centers of nearby
galaxies is close to spherical (e.g., Baganoff et al. 2003; Di Matteo
et al. 2003; Fabbiano et al. 2003). Thus the outer boundary
considered by PB03 capture better RIAF in Sgr~A$^\ast$ than those used
in simulations of high-$l$ flows.

PB03's simulations follow the evolution of the material with 
a range of $l$. The material with $l \gtrsim 2 c R_S$ 
(where $R_S=\frac{2GM}{c^2}$ is Schwarzschild radius)
forms an equatorial torus because
its circularization radius is located outside the last stable
orbit. The torus accretes onto a black hole as a result of
magnetorotational instability (MRI, Balbus \& Hawley 1991). As the
torus accretes, it produces a corona and an outflow which can be strong
enough to prevent accretion of a low-$l$ material (i.e.,
with $l< 2 c R_S$) 
to fall through the polar regions. The torus outflow and the corona 
can narrow or even totally close the polar funnel for the accretion of
low-$l$ material. One of PB03's conclusions is that even a slow
rotational motion of the MHD flow at large radii can significantly
reduce $\MDOT_a$ compared to the Bondi rate.

Many properties of the torus found by PB03 are typical for the  MHD
turbulent torii presented in other global MHD simulations (e.g., Stone
\& Pringle 2001; HB02). Radial density profiles, 
properties of the corona and outflow, as well as rapid variability are
quite similar in spite of different initial and the outer boundary
conditions.

The main difference between  simulations of PB03' and the others is that the former showed 
that the typical torus accretion can be quasi-periodically 
supplemented or even
replaced by a stream-like accretion of the low-$l$ material occurring
outside the torus (e.g., see the bottom right panel of Fig. 2 in PB03). 
When this happens, $\MDOT_a$ sharply increases and then gradually decreases.
The mass-accretion rate  due to this 'off
torus' inflow can be one order of magnitude larger than that from
the torus itself (see Fig.~1 in PB03). 
The off torus accretion is a consequence of
the outer boundary and initial conditions which introduce the low-$l$
material to the system. This material can reach a black hole because
the torus, the corona, and the outflow are not always strong enough to
push it away.

One can expect that in the vicinity of SMBH at the centers of
galaxies, some gas has a very little angular momentum and could be
directly accreted.  Such a situation likely occurs in GC where a
cluster of young, massive stars losing mass surrounds a SMBH
(e.g., Loeb 2004; Moscibrodzka et al. 2006, Rockefeller et al. 2004).
On the other hand, 3D hydrodynamical numerical models 
of wind from the close stars showed that the wind can 
form a cold disk around the Sgr A* with the inner 
radius of $10^3 -10^4$ Schwarzschild radii $R_{S}$ (Cuadra et al. 2006). 
But unless the disk includes the viscosity it can not accrete closer.
In the context of such models we can again expect 
that only a matter with initial relatively low angular momentum,
can inflow at smaller radii at which PB03 started their computations. 
Also the winds from inner 0.5 '' 
stars (SO-2, SO-16, SO-19)
can supply a lower angular momentum matter for the accretion (Loeb 2004).
To test further properties of the low angular momentum scenario, 
we supplement here the dynamical MHD study of PB03 with the computations 
of the time dependent radiation spectra of accreting/outflowing material.

There are only a few papers (HB02; Goldston et al. 2005; Ohsuga et al. 2005)
estimating radiation spectra, or at least the integrated emission,
predicted by MHD simulations of accretion flows
in GC.
These papers assumed different dynamical situations than the one
considered in the PB03 simulations.
HB02 made an
order-of-magnitude estimates of the emission emerging from a MHD torus
in the Galactic Center. In particular, they estimated the peak
frequency of the synchrotron emission coming
from the inner parts of the flow at
$~ 2.5 \times  10^{11}(n_0/10^8)^{1/2}$ Hz 
(where $n_0$ is the scaling density of the ions).
They also calculated a location of the peak of  
a total bremsstrahlung emission 
at about $10^{21}$~Hz.
They concluded that the dynamical variability in the MHD simulations
is in general consistent with the observed variation in Sgr A*.
However, HB02 focused mainly on a dynamical evolution of a high-$l$ 
non-radiative torus and did not consider the detailed
spectral properties of the flow.

Goldston et al. (2005) used numerical simulation of HB02 and estimated
the synchrotron radiation assuming the scaling of the electron
temperature, $T_e \sim T_i$. They calculated the synchrotron part of
the radiation spectrum and modeled polarization variability
including effects of self-absorption.  The emission is variable (by a
factor of 10) at optically thin radio frequencies and originates in the very 
inner parts of the flow (timescales of order of hours). 
The authors predict that the variability at different frequencies should
be strongly correlated. The general conclusion is that
a variable synchrotron emission in Sgr A* could be generated by 
a turbulent magnetized accretion flow.

Ohsuga et al. (2005) presented spectral features predicted by the 3D
MHD flows simulated by Kato et al. (2004). These simulations followed
an evolution of a turbulent torus. 
The heating/cooling balance equation used by Ohsuga et al. (2005) for
calculating the electron temperature includes radiative cooling and
heating via Coulomb collisions.  The authors found that MHD flows in
general overestimate the X-ray emission in Sgr A*, because the
bremsstrahlung radiation originated at large radii dominates the X-ray
band. However, the quiescent state spectrum of Sgr A* cannot be
reconstructed simultaneously in the radio and X-ray bands. 
The authors reconstructed the observed flaring state, with assumption of rather
high $\MDOT_a$ and restrict the size of the emission region
to only 10 $R_{S}$.  A small size of the emission region is
consistent with the short duration of the observed X-ray flares, and it
results in the X-ray variability so there is no need for
additional radiation processes, e.g. radiation from 
nonthermal particles.
On the other hand the observed quiescent X-ray
emission is extended, which contradicts the assumption made in Ohsuga et al. (2005).

Our goal is to compute the spectral signatures of a highly
time-dependent flow found by PB03. In particular, we check whether a
reoccurring switching between various accretion modes present in the
PB03 simulations is consistent with the observed flare activity in Sgr
A*.  Our spectral model assumes that at each moment of the accretion
flow there is a stationary distribution of both hydrodynamical and
magnetic quantities.  Radiative transfer is calculated based on this
'stationary' solution.  The radiation computations are performed
independently from the dynamics (Sec.2). Our results show which parts of the
accretion flow create characteristic features in the radiation spectra. In
particular, we create maps of emission for synchrotron radiation
including self-absorption and Comptonization (Sec.3). We also show time
evolution of the spectra emerging from the flow for a number of moments in
the evolution of an inner torus.  In section 3 we discussed
results in the context of the Galactic Center.
We conclude in Sec.4.

\section{Method}

We computed the synthetic spectra for the simulation denoted as 'run D' 
in PB03. This simulation assumed a slowly rotating
accretion flow (with an angular momentum parameter $l \sim 2 R_S c$ at
the equatorial plane) and were performed in the spherical polar
coordinates.  The computational domain for run D was defined to occupy
the radial range $r_i~=~1.5~R_S \leq r \leq \ r_o~=~ 1.2~R_B$, 
(where $R_B=\frac{GM}{c^2_{\infty}}$ is the Bondi radius, $c_{\infty}$ is a sound speed at infinity ),
 and the angular range $0^\circ \leq \theta \leq 180^\circ$. PB03
considered models with $R'_S\equiv R_S/R_B=10^{-3}$. The $r-\theta$ domain was
discretized into zones with 140 zones in the $r$ direction and 100
zones in the $\theta$ direction.  PB03 fixed the zone size ratios,
$dr_{k+1}/dr_{k}=1.05$, and $d\theta_{l}/d\theta_{l+1} =1.0$ for
$0^\circ \le \theta \le 180^\circ$.

The MHD simulation were performed using dimensionless
variables. Therefore we first rescale all the quantities. The gas
density, $\rho$ is scaled by a multiplication
factor $\rho_{\infty}$  which is the density at infinity.
The magnetic field
scales with $[(B_r^2 + B_{\theta}^2 + B_{\phi}^2 ) \rho_{\infty}]^{1/2}$, 
where $B_r, B_{\theta}$, and $B_{\phi}$
is the radial, latitudinal, and azimuthal component of the magnetic field
respectively.
The internal energy density, $e$ scales with $\rho_{\infty}$ and $c_{\infty}$.

To compute the synchrotron emissivity
we calculate the strenght of the magnetic field, 
$B=(B_r^2 + B_{\theta}^2 + B_{\phi}^2 )^{1/2}$, at
each grid point.

The ion temperature of the gas is calculated using politropic
relation: $T_i= \frac{\mu m_h e(\gamma-1)}{k_b \rho}  $, where
$k_b$, $\mu$, and $m_h$ is the Boltzmann constant, mean particle weight
($\mu=0.5$) and proton mass, respectively.  We set the adiabatic index
$\gamma=5/3$. We use scaling constants appropriate for the case of Sgr
A* i.e. $\rho_{\infty}$ is chosen to fit the observed mass accretion
rate (see Section~\ref{sec:thremal}).

PB03 did not include radiative cooling in their calculations;
therefore, we use $e$ and $\rho$ to compute the distribution of the
temperature of ions $T_i$, using the politropic relation mentioned above. 
However, determination of the electron temperature is the key step in spectral
calculations. Ions and electrons undergo different types
 of cooling and heating processes, and they
are not well coupled in a low density plasma. Ions and electrons
are likely to have different temperatures.
We calculate the electron temperature distribution by solving the
heating-cooling balance at each grid point at a given time in the
simulation. We present a detailed description of the method in Sect.2.1
and outline the radiative transfer method in Sect. 2.2.

\subsection{Heating-cooling balance.}

To calculate the electron temperature in each cell of the grid, we
solve the cooling-heating balance equation for electrons which is given by:
\begin{equation}
Q_{{\rm adv}} = \delta Q_{{\rm pdv}} + Q_{{\rm ie}} - Q_{{\rm cooling}},
\label{eq:1}
\end{equation}
where $Q_{{\rm adv}}$, $Q_{{\rm pdv}}$,$Q_{{\rm ie}}$, and $Q_{{\rm cooling}}$ 
is the advective energy transport, compression heating of the ions, 
ion-electron Coulomb collision term, and radiative cooling, respectively.
The numerical factor, $\delta$ is defined as the fraction of 
the compression energy that directly heats electrons ( $0\leq \delta \leq 0.5$). 

The advection term, $Q_{adv}$ consist of the radial part, shown for example
in Narayan et al. (1998), which can be expressed by:
\begin{eqnarray}
Q_{adv,r}=\frac{\rho k_B v_r }{\mu m_H} \Big{[}\frac{3(3-\beta)}{\beta} + a(T_e)+T_e
 \frac{da(T_e)}{dT_e}\Big{]} \frac{dT_e}{dr} \nonumber
\end{eqnarray}

\begin{eqnarray}
- \frac{v_r k_B T_e}{\beta \mu m_H} \frac{d\rho}{dr}
\end{eqnarray}
in 2D, a term in $\theta$ direction should be added. This is 
a small modification of the radial one:
\begin{eqnarray}
 Q_{adv,\theta} =\frac{\rho k_B  v_{\theta} }{\mu m_H r} \Big{[}\frac{3(3-\beta)}{\beta} +
 a(T_e) +T_e \frac{da(T_e)}{dT_e}\Big{]} \frac{dT_e}{d\theta} \nonumber
\end{eqnarray}
\begin{eqnarray}
- \frac{v_{\theta} k_B T_e}{r \beta \mu m_H } \frac{d\rho}{d\theta},
\end{eqnarray}
where $\beta\equiv P_{gas}/P_{tot}$, $P_{gas}$ is the gas pressure,
$P_{tot}\equiv P_{gas}+P_{mag}$ (we compute the magnetic pressure,
$P_{mag}$ as $B^2/8\pi$).  The plasma parameter $\beta$, $\rho$, $v_r$
radial and $v_{\theta}$ latitudinal velocity, are self
consistently taken from the MHD simulations.  The coefficient,
$a(T_e$) varies from 3/2, for non-relativistic, to 3, for
relativistic electrons.

The compression heating ($Q_{pdv}$ in Eq.~\ref{eq:1}) by the accretion for a 2D accretion
flow, can be written as (M06):
\begin{equation}
Q_{pdv}= \frac{P_{gas}}{\rho} \left( v_r \frac{d \rho}{ dr} + 
\frac{v_{\theta}}{r}  \frac{d \rho}{ d\theta} \right)
\label{eq:2}
\end{equation}
To compute the electron-ion interaction term, $Q_{ie}$,  we use 
a standard formula for heating of electrons by Coulomb collisions
(Stepney $\&$ Guilbert 1983).

The radiative cooling rate, $Q_{cooling}$ in Eq.~\ref{eq:1} includes 
the thermal synchrotron and bremsstrahlung radiation, reduced by 
a mean self-absorption. We compute the self-absorption along fifty
directions. Comptonization terms are
computed both for synchrotron and bremsstrahlung radiation.
As described by Mahadevan et al. (1996) who expressed 
the synchrotron thermal emissivity as a function of $T_e$ with 
three coefficients: $\alpha$, $\beta$, and $\gamma$.
For $T_e= 5 \times 10^8$ K, we used $\alpha=0.08$, $\beta=-10.9$,
$\gamma=9.03$ (there is a misprint in Mahadevan et al.1996
\footnote{Mahadevan et al.(1996), Tab.1 for temperature $T=5 \times
10^{8}$ K should be $\gamma=9.03$}).  For the temperatures lower than
$5 \times 10^8 K$, we model the synchrotron emissivity using the
analytical expression of Petrosian (1981).

We assume that the synchrotron radiation is produced by particles
moving in the mean magnetic field $B$.  
Comptonization cooling terms for both synchrotron and
bremsstrahlung radiation are calculated following the description
given by Esin et al. (1996).  More details and exact formulas for the
radiative cooling are given in M06 (sec. 2.1).

To fully describe the electron temperature $T_e(r,\theta)$, 
we iteratively solve Eq.~\ref{eq:1} including all radial and angular
terms. We consider a very optically thin accretion 
where the heating - cooling balance Eq.~\ref{eq:1} is dominated by two
terms, namely the advection and compression energy rates. These two
rates can lead either to heating or cooling, depending on radial and
angular velocity directions, density derivatives, and $\frac{dT_e}{dr}$
or $\frac{dT_e}{d\theta}$ sign.  $Q_{ie}$ always heats electrons, while 
$Q_{cooling}$ always cools electrons.  If $Q_{adv}$ and
$Q_{pdv}$ are negative, (in a sense that $Q_{pdv}$ and
$Q_{adv}$ are both cooling terms ), one can find a solution of the balance
equation by decreasing electron temperature.
This means that we are taking more and more energy from ions through
Coulomb collisions, to keep electrons balanced. Nominally in the ion
equation of the conservation of energy also the Coulomb coupling term
should be included. This process can make electrons indirectly 
dynamically important even if $T_e << T_i$.
In our calculations,
$Q_{ie}$ is small compared to $Q_{pdv}$ so that electrons do
not play role in dynamics.  If, on the other hand, $Q_{adv}$ and
$Q_{pdv}$ are positive, (in a sense that $Q_{pdv}$ and
$Q_{adv}$ are both heating terms), radiative processes 
couldn't cool the electrons efficiently to keep
an energy balance.  These are two extreme cases that are quite
difficult to treat numerically (especially the second case, where
the heating energy cannot be balanced by the radiative cooling
within above described framework).

However these issues can be avoided by including
additional cooling processes (for instance creation of electron
positron pairs) or, as we have found, by smoothing velocity and
density radial profiles at each angle.  We adopted the later and we
used linear interpolations for the velocity and density radial
profiles, over the whole grid.  This operation 
results in the advection and compression rates being always negative and
positive, respectively.  The equation of energy can be then solved
using the standard formula for cooling mechanisms.

To avoid the same problem in $\theta$ direction, we neglect all
angular terms in $Q_{adv}$ and $Q_{pdv}$. The energy balance
equation, that we use, reduces to:
\begin{eqnarray}
  \frac{\rho k_B v_r }{\mu m_H} \Big{[} \frac{3(3-\beta)}{\beta}
 + a(T_e)+T_e \frac{da(T_e)}{dT_e}\Big{]} \frac{dT_e}{dr}\nonumber
\end{eqnarray}
\begin{eqnarray}
-  \frac{v_r k_B T_e}{\beta \mu m_H} \frac{d\rho}{dr} 
=\delta \frac{P_{gas}}{\rho} (v_r \frac{d\rho}{dr}) \nonumber
\end{eqnarray}
\begin{eqnarray}
 + Q_{ie} - Q_{cooling}(T_{e,initial})
\label{eq:reduced}
\end{eqnarray}
Omitting the terms in the $\theta$ direction does not significantly
affect $T_e$ calculations in the regions where $|v_\theta/v_r|<1$.
We find that this is the case in the equatorial region where $|v_\theta/v_r|<1$.
However,  more sophisticate $T_e$  
calculations would have to be carried out for the regions 
where PB03b's simulations show a significant torus outflow.
We emphasize that we use the original
values of $\rho$ in calculating spectra but used smoothed $\rho$ and $v_r$ only
 in order to calculate $T_e$.  We note
that our method of calculating the  heating-cooling balance assumes a
stationary accretion at each time step the MHD simulation.  Actually,
the flow is strongly turbulent, the energy advected in the angular
direction could return to the same point at some later time, and the
net heating-cooling balance in $\theta$ direction can be close to
zero.

An initial value of the electron temperature in each cell of the grid
is given by the adiabatic relation $T_e(i,j) =
(T_{e,\infty}/(\rho_{\infty})^{\gamma-1} \rho(i,j)^{\gamma-1}$.
In the initial loop we calculate the cooling term $Q_{cooling}$ at
each point, and solve Eq.~\ref{eq:1} in each point. After the first
iteration we correct $T_e$ in the whole simulation domain, and begin the
procedure again.
 We stop the iterations when the equilibrium is
reached in each zone. The equilibrium condition
requires that the electron temperature at a given place
does not change in the next iteration very much, we usually obtain
few percent consistency.  

Two most dominant terms in  Eq.~\ref{eq:reduced} are $Q_{adv}$ and
$Q_{pdv}$, thus $\delta$ parameter cannot be very large.  We
show this in the case of radial equations. When we approximate
gradients of electron temperature and density with the 
algebraic values (Eq. $dT_e/dr=T_e/r$), we obtain:

\begin{equation}
\frac{Q_{adv}}{ \delta Q_{pdv}} \sim \frac {T_e}{\delta T_i}
\end{equation}

Here we assume $\delta=0.01$ in our calculations.  If the radiative cooling is
neglected, then $T_e \sim \delta T_i$. In our case the radiative
cooling, calculated at the initial grid of temperatures can indeed be very
small, partly due to a weak magnetic field. In the PB03's simulation,
$\beta \sim$ 1 in most of the zones ($\beta > 0.9$ is in $65\%$ of the
grid points).

Our calculations show that the ions can reach very high temperatures of
$10^{12} K$ in some regions: such temperatures are too high for electrons. 
Therefore, to have
$T_e \sim \delta T_i$ in order to keep the model self-consistent, we
assume rather low values of the parameter $\delta$. Our analysis is
consistent with Quataert $\&$ Gruzinov's (1999) conclusions that for
weakly magnetized plasma $\delta$ should be rather small. We keep
$\delta$ constant in the entire flow, as a global parameter. This
approach should be refined in future studies, because $\beta$ 
varies across the computation domain, in particular it is small
very close to the black hole.

With the smoothed $v_r$ and $\rho$ distribution, the values of
higher $\delta$ would lead to the higher values of $T_e$, and an
additional cooling process would be needed.  We suppose that a creation
of electron-positron pairs could be produced and start to dominate the
cooling term.  
Currently the maximum value of $T_e$ in our
calculations is a few $\times 10^{10}$~ K in the innermost regions.
For higher temperatures, such as $10^{11} K$, and the assumed electron
positron pair equilibrium, the ratio between the pair number density
$n_{+}$ and the number density of protons $n_p$, $z$ steeply rises
from z=$10^{-4}$ up to z=16 in the inner parts of the flow.

The effects discussed above could change if we increase the density scaling
factor, $\rho_{\infty}$.  The bremsstrahlung emissivity increases like
$\rho^2$, so the advection or compression term  could be more
easily balanced by the radiation cooling.  These results are thus
correct only for very optically thin accretion flows ( Thomson
thickness is $\tau_{th}<<$ 1), where cooling by radiation is
negligible as far the flow dynamics is concerned.

\subsection{Monte Carlo calculations of radiative transfer.}

Our Monte Carlo algorithms are adopted from Pozdnyakov et al.  (1983)
and Gorecki $\&$ Wilczewski (1984) descriptions.  Distribution of
photon emissivity and the distribution of photon energy are calculated
as described in Kurpiewski $\&$ Jaroszynski (1999).  We consider a 2-D
flow and the place of emission is given by the conditional
probability.  We take into account a 2D distribution of the rate of
$\dot{N}_{ij}$ photon emission (for details how do compute $\dot{N}_{i,j}$ 
see Kurpiewski $\&$ Jaroszynski 1999).
When choosing a random location of the photon emission, we calculate 
the probability of  the emission at a given radius: 
\begin{equation}
P(r_i)=\frac{\sum_{j} \dot{N}_{ij} } {\sum_{i,j} \dot{N}_{ij} },
\end{equation}
where i and j denote the radius, $r$,  and the angle, $\theta$, respectively.
If the radius of the emission is chosen, we determine the angle of the emission from
the distribution of angles at a given radius using formula:

\begin{equation}
P(r_i,\theta_j)=\frac{ \dot{N}_{i,j}} {\sum_{j} \dot{N}_{i,j}}.
\end{equation}

The only difference here in the comparison to method of Kurpiewski
$\&$ Jaroszynski (1999) is that we calculate the self-absorption of
photons
along the line of sight (the viewing angle of the observer 
is one of the free parameters).

Finally we note that to calculate the radiation spectra, we follow M06
in using a grid which consists of 100 zones in the radial direction and 100
zones in $\theta$ direction.
Therefore, we transform all the input data: ion
temperature, magnetic field induction and density taken 
from the PB03 simulation, and calculated electron temperature,
 into a new grid as defined in M06.

\section{Results}

\subsection{Models with radiation from thermal particles}\label{sec:thremal}

\label{sect:basic}

In general, the broad band spectrum of a hot, optically thin flow
consist of three components: a synchrotron bump extending from the
radio to IR band, a Comptonization component at higher energies than
the synchrotron bump, and a bremsstrahlung bump extending from the
IR-UV to X-ray and $\gamma$-ray energies. Specific radiation output
emerging from a given flow depends 
not only  on the gross properties of the flow such as - the mass
accretion rate onto a SMBH - but also the flow and magnetic field structure.
 
The input parameter is $\rho_{\infty}$.
We choose value of this parameter that are
suitable for the conditions in Low Luminosity Active
Galactic Nuclei (LLAGN). In particular, our basic model is calculated
for the case of the Galactic Center. We assume a black hole mass of $3.7
\times 10^6$ $\rm{M_{\odot}}$ and the asymptotic value of the density
of $6 \times 10^{-20}$ $\rm{g/cm^3}$.
  
To determine which characteristic flow patterns are
responsible for specific spectral properties we calculated broadband
radiation spectra for 32 snapshots from the MHD simulation, including the
four characteristic accretion states, A, B, C, and D, discussed by PB03.

State A is an accretion state at the early phase of the simulations,
while state B and C corresponds to an accreting and non-accreting torus, 
respectively. Finally, state D corresponds to 
a stream-like accretion of the very low-$l$ material 
which is approaching the SMBH not through the torus but from outside of
the torus (see the bottom-right panel 
of  Fig. 8 in PB03b where the stream
is below the equatorial plane).
The four characteristic states differ not only in the flow and magnetic
structure but also in their gross properties, e.g., $\MDOT_a$.
In particular, $\dot{M}_a/\dot{M}_{B}$=0.11, 0.009, 0.001,
and 0.048, for state A, B, C, and D, respectively ($\dot{M}_{B}$
is the Bondi accretion rate).

To study time variation of the radiative properties, in particular
to construct the light curves for various wavelengths, we compute radiation 
spectra not only for these four states but also
for another 28 snapshots, for the time, $t$
from 2.35 to 2.40
(as in PB03, all times here are in units of the Keplerian orbital 
time at $r=R_B$). Here, we include only thermal
particles, and assume the inclination
angle i=$90^\circ$ (edge on view).

\begin{figure*}
\includegraphics[width=1.0\textwidth]{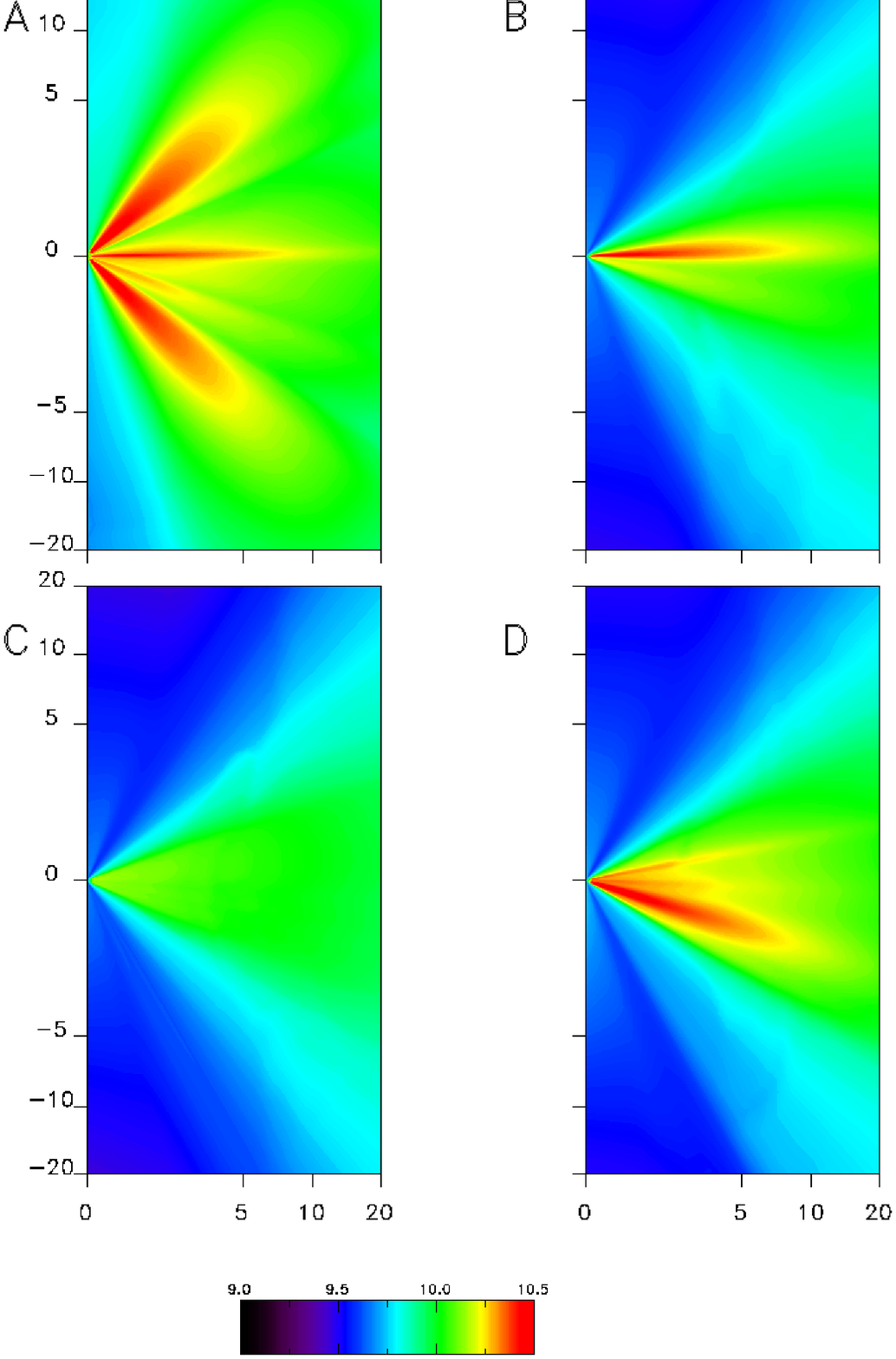}
\caption{ 2-D color maps of electron temperature
for four characteristic accretion sates
A,B,C,and D. 
Length scale for the horizontal axies is logarithmic 
in units of $R_{in}$, so that the size of a box is 20 $R_{S}$.
The temperature is given in logarythmic scale in K.
}
\label{fig:0}
\end{figure*}

 We begin presentation of our results with showing maps of $T_e$ for the 
four characteristic states (Fig.~\ref{fig:0}).
The figure shows only the innermost 20~$R_S$ 
where the temperature ranges between $\approx 3 \times 10^9$~K
and $\approx 3 \times 10^10$~K. The temperature distribution traces somewhat 
the density distribution but there is no one-to-one correspondence
with the density (see Fig. 8, in PB03b) 
because $T_e$ is a non-linear function of a few quantities,
including the complex velocity field. 
In addition, the temperature maps compared to the density maps, 
lack fine details due to our smoothing procedure described in \S2.1.
The maximum electron temperature in our models is larger
than the maximum temperature in ADAF models. For example,
Quataert \& Narayna's (1999) calculations
showed that to obtain $T_e$ as high as in our calculations one should 
assume $\beta=10$ and a relatively strong outflow
(e.g., model 5b in their table 2).

\begin{figure*}

\includegraphics[width=1.0\textwidth]{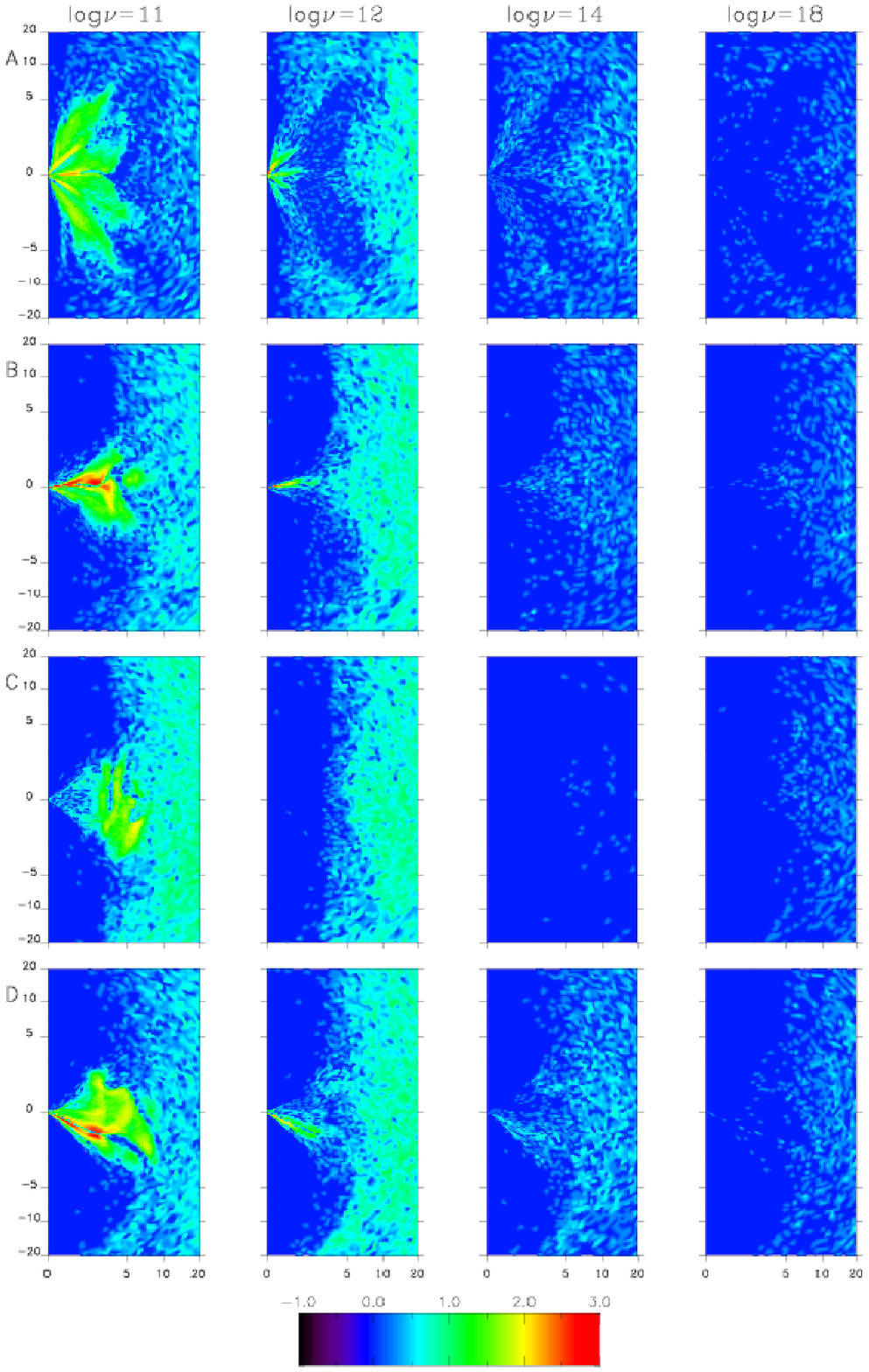}

\caption{ 2-D color maps of the emission 
at four frequencies: $\log \nu=$11,12,14, and 18 Hz (from left
to the right) and for four distinct accretion sates
A,B,C,and D (from top to
bottom; see the main text for more detail on state A,B,C, and D). 
Length scale for the horizontal axies is logarithmic 
in units of $R_{in}$, so that the size of a box is 20 $R_{S}$.
}

\label{fig:1}

\end{figure*}

Fig.~\ref{fig:1} shows the maps of synchrotron emission for the four
characteristic states and for four frequencies; $\log \nu=11,12,14,18$.
At these frequencies emission is dominated by synchrotron emission ($\log \nu=11 \& 12$),
Comptonization ($\log \nu=14$) and bremsstrahlung ($\log \nu=18$).
The figure shows only the innermost 20~$R_S$ 
where most of the synchrotron emission is produced.

First, we describe our results for synchrotron emission (two left columns in Fig.~\ref{fig:1}).
In the state A, the photon emission is distributed over a
wide range of $\theta$. In state B, the majority of photons
come from the very inner cusp of the accretion torus. In state C,
there is no cusp, because the torus is truncated by the magnetic field, 
and the synchrotron emissivity  is reduced in the very inner region. In 
state D, the emissivity is relatively high and most of
the synchrotron photons are created in the stream of matter
approaching the black hole from below the equatorial plane.

As expected, the synchrotron photons undergo Comptonization.
For states B, C, and D, relatively few scatterings 
occur near the black hole (see the third column from left).
Generally, the scattering regions are similar in all four states, except 
for the enhanced Comptonization in a narrow elongated region below 
the equator for 
$\log r<0.5$ in state D. This narrow region corresponds to 
the stream of low-$l$ matter mentioned above.
For state A, scattering is more uniform compared to the above three states 
because the density distribution is also more uniform. 

In all four states the bremsstrahlung photons are created relatively far
from the center (i.e., $90\%$ of photons is created beyond $50 R_{S}$, 
see the right most column). 
In the central part of the accretion flow, the distribution of 
the bremsstrahlung emission has a cusp for all states but state C.
Because the bremsstrahlung emissivity traces the density distribution,
the further from the center, the more uniform
bremsstrahlung emissivity.
Only the highest energy bremsstrahlung photons are created close to 
the black hole. 

\begin{figure*}

\begin{center}

\includegraphics[width=0.8\textwidth,angle=90]{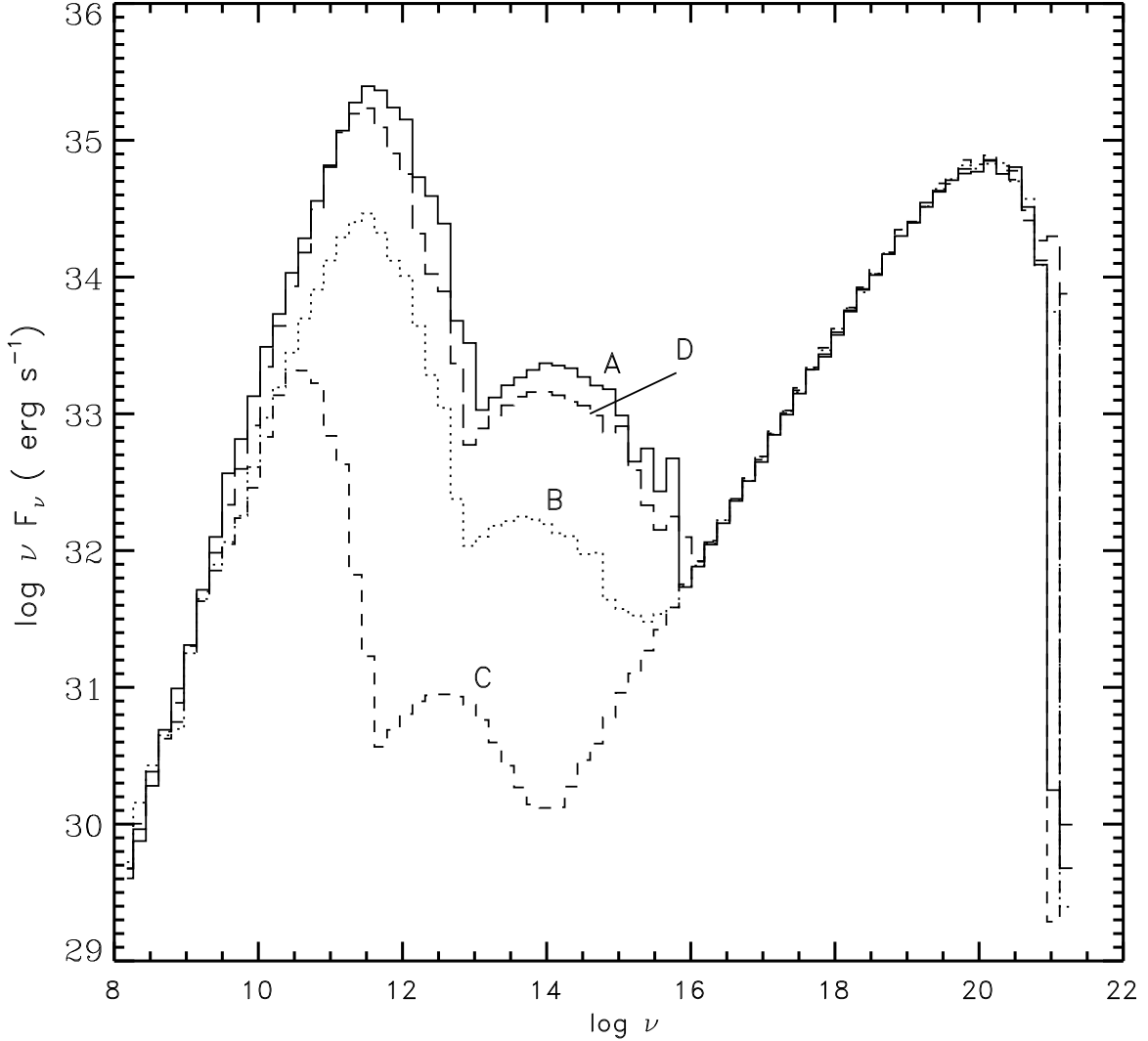}

\caption{ Radiation spectra 
corresponding to accretion states A, B, C, and D, respectively
(only thermal electrons were included). The  mass accretion
 rate at the outer boundary of the simulation, is 
$\dot{M}_{B}= 3.7 \times 10^{-6} \rm{M_{\odot}/yr}$ and 
central black hole mass is $M_{BH}=3.7 \times 10^{6} \rm{M_{\odot}}$.
In state A, mass accretion rate (through the inner boundary of the flow) 
in units of $\dot{M}_{B}$, 
equals about 0.11, in the
state B is $\sim$ 0.009, in the state C, when the torus does not
accrete, it decreases to 0.001. For the state D 
$\MDOT_a$ increases to 0.048.}

\label{fig:2}

\end{center}

\end{figure*}

Fig.~\ref{fig:2} shows the radiation spectra for the four accretion states. 
All spectra have three distinct components:
the synchrotron, Compton and bremsstrahlung bump at low,
medium, and high frequencies, respectively. The spectra differ
from each other mainly in the location and strength of the synchrotron
and Compton bumps. The changes in the synchrotron radiation
reflects the fact this radiation
is produced in the central part of the flow where the flow changes 
the most from state to state. In particular, 
the low ($10^8-10^{11}$ Hz)
and high ($10^{11}-10^{13}$ Hz) frequency synchrotron radiation changes
because this radiation is produced respectively in  the thick torus
and in the plunging region.
Comptonization of synchrotron photons occurs predominantly in the
outer regions of the torus. However, the Comptonization bump
($\nu = 10^{13}-10^{15}$ Hz) changes from state to state mainly  because
of the changes of the energy of the input synchrotron photons. 
On the other hand, most bremsstrahlung emission
($\nu = 10^{15}-10^{21}$ Hz) is produced in the outer regions where
all four states are quite similar.
Therefore this part of the spectrum does not change, except for the high energy
cutoff which is sensitive to the conditions in the inner flow 
where the highest energy bremsstrahlung photons are produced there.

Cross comparison between Figs. 1 and 2 shows that there is no
one-to-one correspondence between the accretion state 
and the spectrum. For example, states A and D are quite different
(including their corresponding $\dot{M}_a$) yet their predicted
spectra are relatively similar. 
However, it is not
to say that radiation spectra do not depend on the the accretion state
because two very different states: C and D have very different spectra.
In particular,
the flux of the higher energy  
synchrotron radiation and Compton radiation is orders of magnitude
higher for the stream accreting state D than for the suppressed accretion state C.
Similarly, the emissivity is also significantly 
higher for the torus accreting state B, than for state C.

We find that the radiation spectrum responds to changes in 
the accretion state and mass accretion rate in a complex non-linear way.
Because the flow is asymmetric the spectrum can also depend on 
the inclination angle, $i$.

To check the spectrum dependence on $i$, we compute several spectra for 
various inclinations. 
The emerging radiation spectrum is relatively insensitive to $i$, which is
consistent with our optically thin approximation (i.e.,  low
gas density and column density).
However, even for these conditions self-absorption can be
appreciable for the synchrotron radiation because the opacity
at low frequencies ($10^8-10^{11}$ Hz) is sensitive to 
the electron temperature, the density  and the magnetic field distribution 
function (which depends on emissivity function). 
At higher frequencies (higher than the synchrotron peak frequency) 
the spectra are independent of $i$, because
self-absorption does not play a role in this part of the spectrum.

\begin{figure*}

\begin{center}

\includegraphics[width=0.8\textwidth,angle=90]{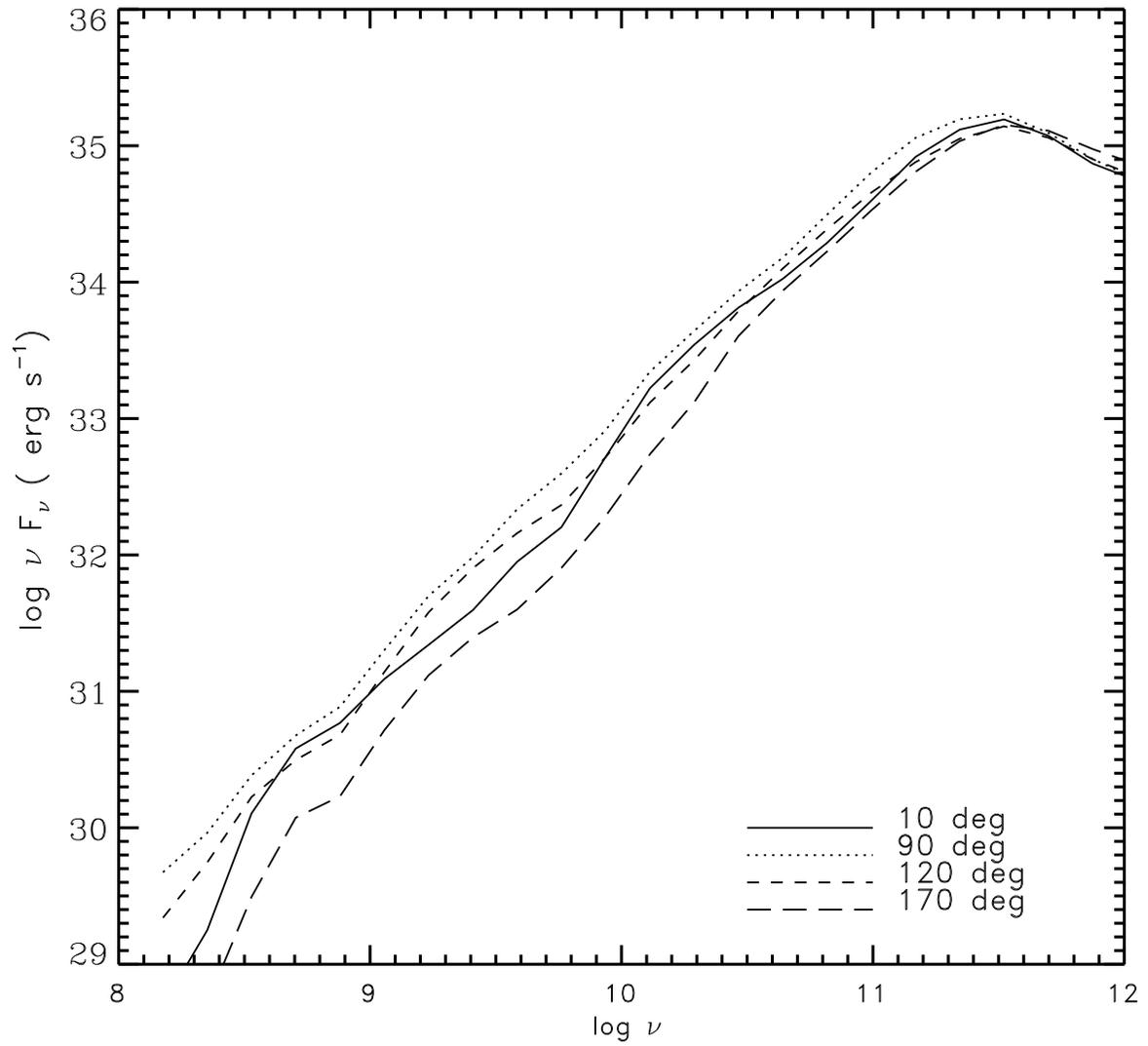}

\caption{ Synchrotron spectra emitted at the accretion
state D in the radio band, seen by observers at the different locations.}

\label{fig:3b}

\end{center}

\end{figure*}

Fig.~\ref{fig:3b} shows our results for four inclination
angles: $i=$ $10^{\circ}, 90^{\circ}, 120^{\circ}, \& 170^{\circ}$.
The spectra were computed for state D for which
effects of $i$ are the strongest as this state 
represents the episode of the highest flow asymmetry due to the 
presence of the stream of the low-$l$ material below the equator.
We present results for  $\nu < 10^{12}$ Hz because
for high $\nu$ effects of `i' are very small as discussed above.

The largest spectral differences are 
for $i=90^{\circ}$ and $i=170^{\circ}$: the flux can be an order on magnitude
lower for the former compared to the latter.
For  $i=90^{\circ}$, an observer can see low energy photons
produced in the outer parts of the flow whereas for 
$i=170^{\circ}$, low energy photons are self-absorbed 
in the dense accreting material. 
For $i=120^{\circ}$, an observer sees the center through the accreting
material, the self-absorption is somewhat stronger, and the emission
decreases at low synchrotron energies.
We conclude that the spectra weakly depend on $i$ and it will be very difficult
to infer the inclination angli based on spectral analysis.

\begin{figure*}

\begin{center}
\includegraphics[width=0.8\textwidth,angle=90]{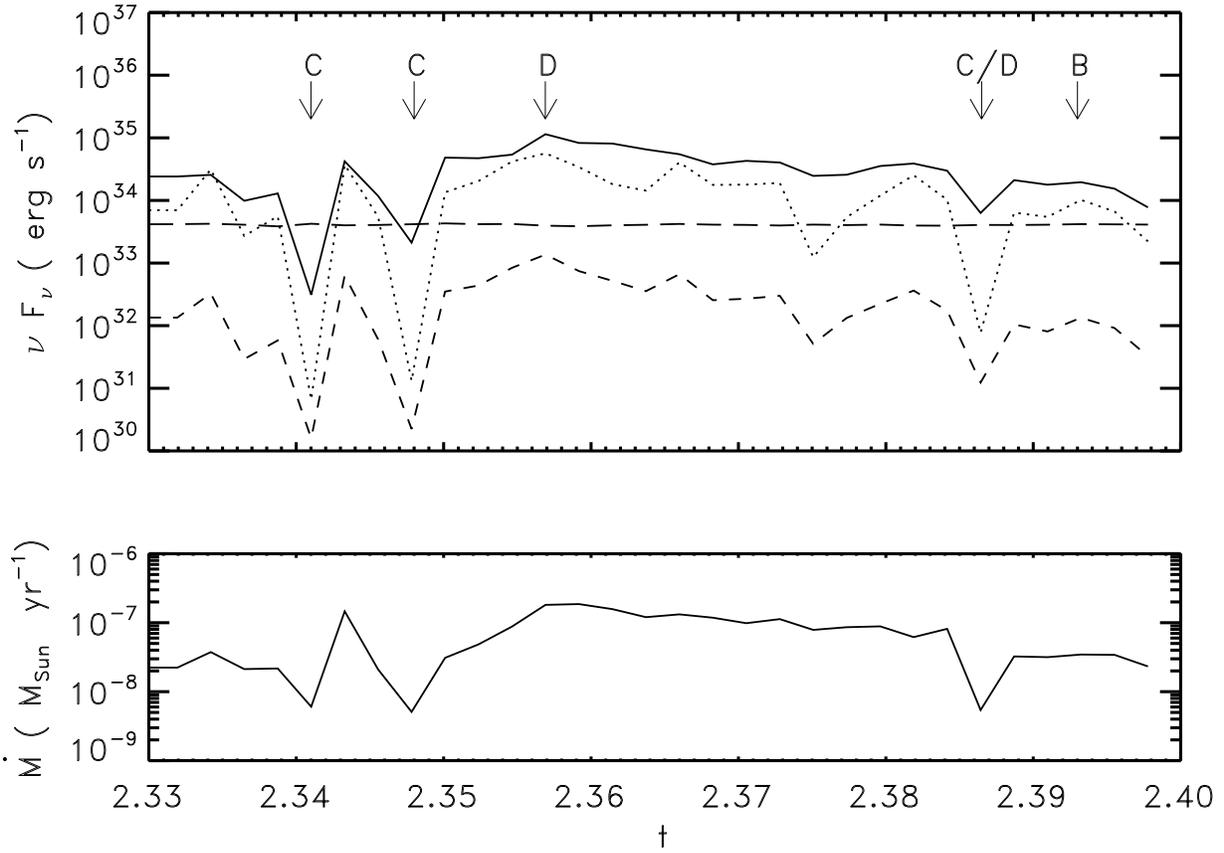}

\caption{Upper panel: Time evolution of 31 spectra in time interval from 2.33 to 2.4.
Time is given in Keplerian orbital time (see PB03).
$\rm{t_{Keplerian}=2 \pi R_{B}/ \sqrt{GM/R_{B}}} = 10^7$ seconds, where
$\rm{R_{B}=10^3 R_g}$. 
 Each curve represents a monochromatic
flux at different frequencies: $10^{11}$ (solid),  $10^{12}$ (long dash), 
$10^{14}$ (short dash) and $10^{18}$ Hz (dotted line). X-ray ($10^{18}$ Hz)
variability is very low in comparison to the variability at the other
wavelengths. Observations at $10^{11}-10^{14}$ Hz can be used to distinguish
the characteristic accretion states B $\&$ D. 
Lower panel: the time dependence of the mass accretion rate for comparison.}

\label{fig:4}

\end{center}

\end{figure*}

Fig.~\ref{fig:4} shows light curves at four frequencies from radio to X-rays 
(the top panel) and the corresponding time evolution of $\MDOT_a$ 
(the bottom panel).
In the top panel, solid, long dashed, short dashed, and dotted line corresponds
to the flux at $\nu = 10^{11}$, $10^{12}$, $10^{14}$, and $10^{18}$ Hz, respectively.
We analyze the time behavior for  $2.33 \le t \le 2.40$. 
This time interval is representative of the late-time evolution of 
the flow as it captures flow switching between three of 
the four characteristic states: B, C, and D (state A is characteristic
only of the early-time evolution). 
For example, at $t=2.368$ and 2.394, the flow
accretes onto the SMBH through a stream of low-$l$ material (state D)
and a torus (state B), respectively. These accretion episodes were 
described and discussed in detail by PB03. Additionally, 
our analysis includes times (i.e., at $t=2.341$ and 2.348) when
the accretion is suppressed.
During these two episodes the flow is in a state similar to 
state C using PB03's terms.
Moreover, at $t=2.388$ the flow is in 
a transition
between state D and short-lived state C. 

As one might expect based on the spectra presented in Fig.~2,
the strongest variability, up to 3 orders of magnitude, is at
$\nu = 10^{12}$ and $10^{14}$ Hz.  
A smaller variability, yet up to 2 orders of magnitude, is 
in the synchrotron radiation at  $\nu = 10^{11}$ Hz.
Whereas there is practically no variability in the X-ray flux ($\nu=10^{18}$ Hz).
The X-ray flux ratio for different states varies by a factor of 1.06.

The light curves in the radio and IR frequencies show that the response of 
the radiation properties to the changes in the flow dynamics
is the strongest  when there is a significant temporary decrease in $\MDOT_a$
(by nearly two orders of magnitude)
corresponding to a suppression of SMBH accretion 
(i.e., at $t=2.341$, 2.348, and 2.388).
However, even during these changes, the X-ray light remains constant
because, as we described above, bremsstrahlung photons are
produced far away from the center where the flow changes are small.
In summary, the accretion states of  the inner flow cannot be identified in 
the X-ray energy band but they can be identified in the radio and IR bands.

\subsection{Models with radiation from nonthermal particles}

 To model a nonthermal population of electrons
with modified synchrotron emissivities, we follow
Yuan et al. (2003). For simplicity, we neglect cooling
break in nonthermal distribution of electron velocities.
The nonthermal power-law electron distribution,
as a function of Lorenz factor $\gamma$, is given by: 
\begin{eqnarray}
  n_{pl}(\gamma)=N_{pl} (p-1) \gamma^{-p},   ~~~~~~       \gamma_{min}<\gamma<\gamma_{max}
\end{eqnarray}
where $N_{pl}$ is a total number of nonthermal particles, $p$
is a power law index and $\gamma_{\rm{max}}$ and $\gamma_{\rm{min}}$ is 
the maximal and minimal Lorenz factor of 
nonthermal electrons respectively. 
The nonthermal and thermal electron distribution functions
connect smoothly at $\gamma_{\rm{min}}$ [i.e. $n_{th}(\gamma_{\rm{min}})=n_{pl}(\gamma_{\rm{min}})]$,
where $n_{th}(\gamma)$ is a thermal distribution so that we are able to calculate
 $\gamma_{\rm{min}}$ and $N_{pl}$ selfconsistently within the model (Yuan et al. 2003).

We adopt the same electron temperature distribution 
as in the thermal case before, i.e. we do not include nonthermal emission in
cooling processes while computing $T_e$.
We assume that $\eta$, which is nonthermal to thermal energy ratio, 
is constant. Parameter $\eta$ is one of the three free 
parameters in calculation of nonthermal radiation. 
Other two free parameters are $\gamma_{max}$ and $p$.
The maximum of a nonthermal emission is at the critical frequency of
$\nu_c=1.5 \gamma_{\rm{max}}^2 \nu_b$, where $\nu_b$ is the cyclotron
frequency (Mahadevan $\&$ Quataert 1997).

To compare nonthermal spectra we assumed $p=2.5$.
This choice of $p$ is motivated by simulations of particle acceleration in
non relativistic and relativistic shocks which showed that $p$ should be
slightly larger than 2, (e.g., Kirk $\&$ Schneider 1987).  
In a nonthermal case, we assumed $\eta$ to be 
very low, $0.1\%$, which was motivated by the work of 
Ghisellini et al. (1998) who showed that self-absorption plays the main role in
thermalization of nonthermal electrons in optically thin hot sources.
We computed the emerging radiation spectra
for $\gamma_{\rm{max}}$ = $10^5$ and $10^6$. These rather high 
$\gamma_{\rm{max}}$ allows for a strong nonthermal emission
at X-rays.

\begin{figure*}

\begin{center}
\includegraphics[width=0.8\textwidth,angle=90]{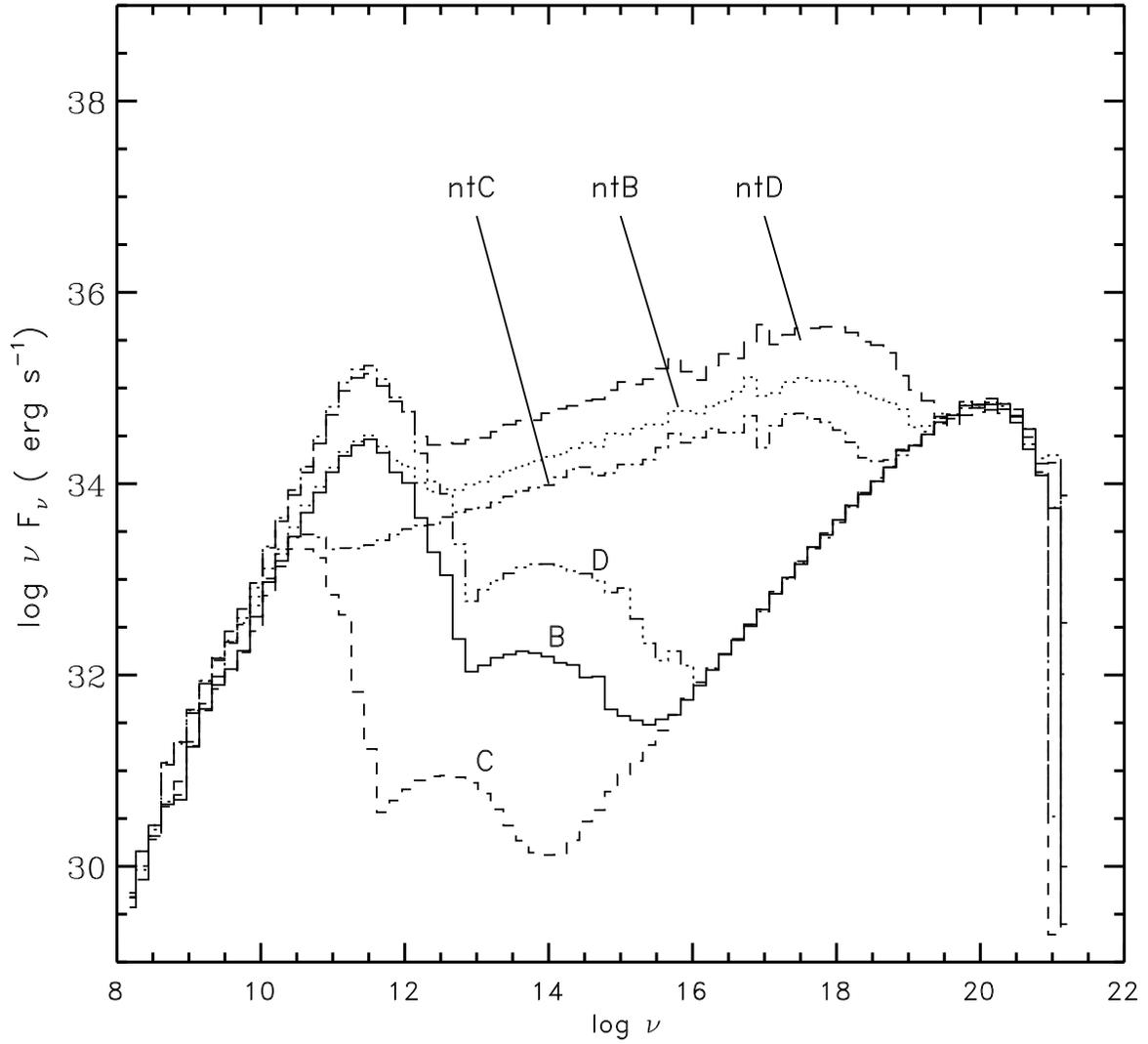}

\caption{ Spectra emitted at the accretion states
 B, C, and D. The model was calculated for the
parameters of Sgr A*. Spectra for thermal electrons are marked by B, C, D, 
and hybrid electron distribution including non-thermal electrons is marked
by `nt B', `nt C', and `nt D'. The parameters for non-thermal electrons are:
$\eta=0.1\%$, p=2.5, $\gamma_{max}=10^{5}$.}

\label{fig:6}

\end{center}

\end{figure*}

\begin{figure*}

\begin{center}
\includegraphics[width=0.8\textwidth,angle=90]{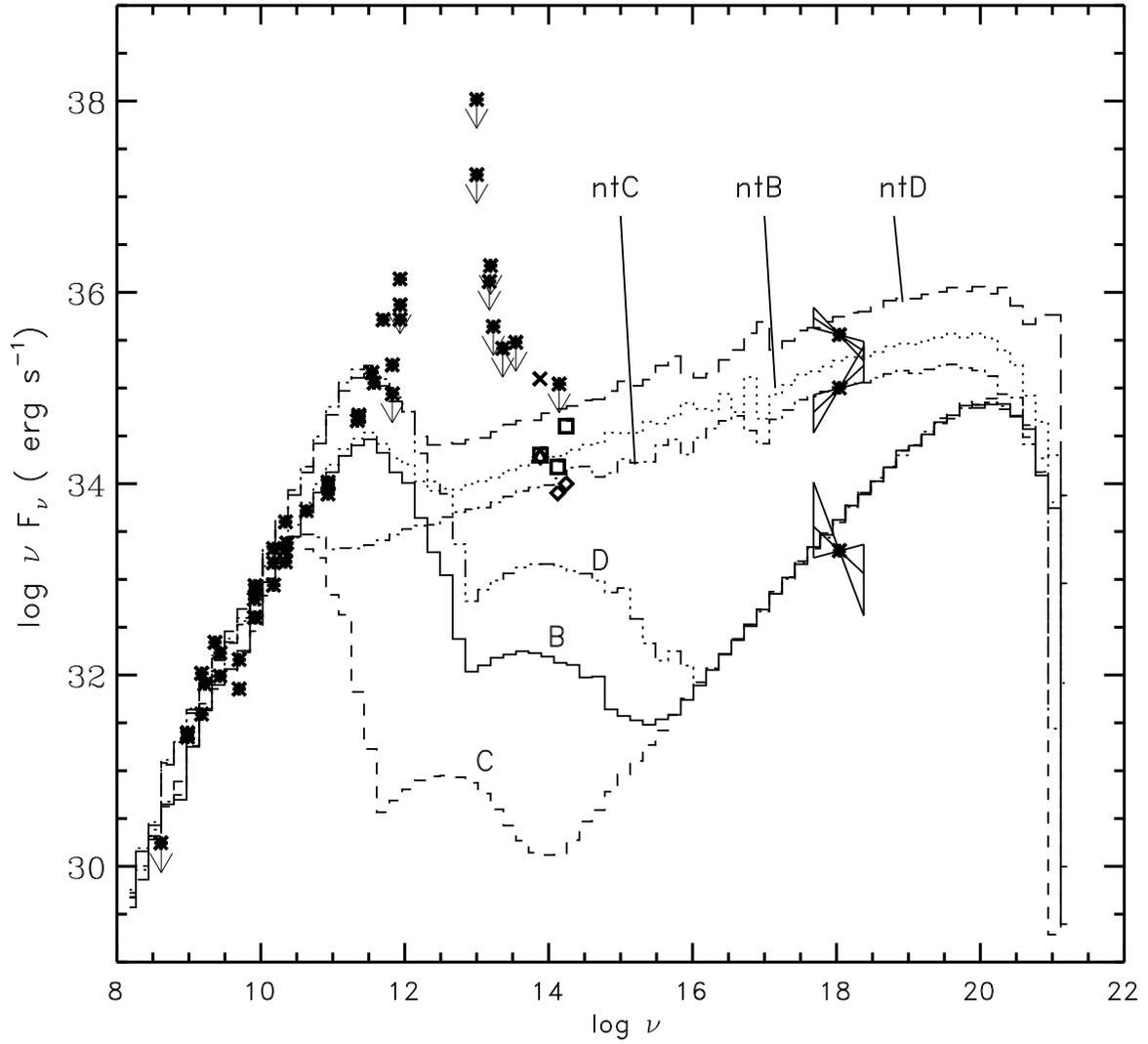}
\caption{ As in Fig.6 for different parameters of non-thermal electron distribution.
Note the change in the non-thermal spectra. 
The parameters for non-thermal electrons are:
$\eta=0.1\%$, p=2.5, $\gamma_{max}=10^{6}$.
The observational data are taken from: Narayan et al.(1998), Genzel et
al.(2003), Ghez et al.(2004), Baganoff et al.(2003), Porquet et
al.(2003).}
\label{fig:6a}

\end{center}

\end{figure*}

Figs.~\ref{fig:6} and~\ref{fig:6a} compare spectra computed  for B, C, 
and D accretion states, with and without nonthermal electrons. 
Nonthermal electrons produce a power law spectrum extending from
thermal synchrotron peak up to the bremsstrahlung bump, or even up to
the $\gamma$-ray energy range for some parameters. 
The IR emission due to nonthermal electrons is much larger in the case with nonthermal
emission, than in the case of pure thermal radiation.
The synchrotron
thermal Comptonization bump cannot be observed in any of the examples,
because it is `hidden' under nonthermal photons. 

For fixed $p$, $\eta$, and $\gamma_{\rm{max}}$ parameters, changes between
accretion states B, C, and D naturally produce variability in the whole
range of frequencies (see below for a discussion of this in
reference to Galactic Center flaring behavior). We note however that
a significant variability in the high energy part of the spectrum
and spectral slopes of the power-law emission
can also be caused by changes in these three free parameters for 
a given accretion state. Changes in $p$ affect the spectral 
slope of the nonthermal spectrum, whereas changes in
$\eta$ affect a nonthermal emission in the whole spectral
range. Generally the flux increases with increasing $\eta$ at all frequencies. If $\eta$
is high enough, no thermal synchrotron peak can be seen in the
radiation spectra, because the whole radiation spectrum is then dominated
by a nonthermal emission.  The position of the high energy cutoff of
the nonthermal emission depends on $\gamma_{\rm{max}}$.
If $\gamma_{\rm{max}}$ and $\eta$ parameters are high
enough, the thermal bremsstrahlung
peak in the high energy part of the spectrum will be covered by nonthermal emission. 
We also note that timescales of the
variability of nonthermal emission in various accretion states can be
very short because nonthermal electrons are created much closer to
the black hole horizon.  This effect could lead to a timescale
variability in high energy bands when the flow changes between different accretion states.

\section{Discussion}

Here, we computed synthetic spectra based on MHD
simulations of an accretion flow performed by PB03.
Using Monte Carlo techniques, we took
into account synchrotron, bremsstrahlung and Comptonization
processes. 
We studied time dependent radiative properties of MHD
flow by investigating light curves and broad band spectra and
effects of inclination.
We also studied effects of a possible non-thermal
contribution to the thermalized electrons.
We discuss  our results in the context the
observation of Sgr A*.

\subsection{Sgr A*- model vs. observations}

As described in Sec. 1 Sgr A* is an underluminous and variable source.
For example, in the X-ray band, multiple flares (e.g. Eckart et al. 2004, Belanger
et al. 2005) are superimposed on a steady, extended emission at the
level of $\sim 2.2 \times 10^{33}$ erg s$^{-1}$ cm$^{-2}$, with
occasional extremely bright eruptions [Baganoff et al. 2001, Porquet
et al. 2003; maximum flux of $1.0~(\pm 0.1) \times 10^{35}$ and
$3.6^{+0.3}_{-0.4} \times 10^{35}$ ergs/s, respectively]. The duration of the
flares ranges from half an hour to several hours, while the rise/decay
time is found to be of the order of a few hundred seconds (Baganoff
2003).

A variable emission is also seen in the NIR band (Genzel et al.  2003,
Ghez et al. 2004).  In two of the events, a 17 min periodicity was
found (Genzel et al. 2003), and recently confirmed by Eckart et
al. (2006a).  X-ray and NIR outbursts are directly related, as shown by
the detection of simultaneous NIR/X-ray events (Eckart et al. 2004,
2006b).  The duration of events is of order of tens of
minutes. Quiescence emission is at the level of $\sim 1.9$ mJy (Eckart
et al. 2004).

For $\nu < 43$ GHz, and $\nu > 10^3$ GHz the observed spectral index in Sgr A* 
is $\alpha_{sp}=0.2$ ($F_{\nu} \sim \nu^{\alpha_{sp}}$), and
$\alpha_{sp}=0.8$  respectively (Narayan et
al. 1998, Yuan et al. 2003 and references there in).  In the X-ray
band, the quiescent emission is fitted well by the power law photon index
$\Gamma=2.7^{+1.3}_{-0.9}$ ($\nu F_{\nu} \sim \nu^{2-\Gamma}$), or it
can be explained by an absorbed optically thin thermal plasma with
$kT=1.9$ keV.  In the intermediate and strongest flare state
$\Gamma=1.3^{+0.5}_{-0.6}$;$\Gamma=2.5^{+0.3}_{-0.3}$ (Baganoff et al. 2001, Baganoff et
al. 2003, Porquet et al. 2003).

\subsubsection{Properties of broadband spectra}
Fig.~\ref{fig:6a} shows the available observational data
for Sgr A*. 
For pure thermal particles, the predicted emission is definitely too
faint in most of the energy bands.
Our models reproduce the level and slope of the radio emission for low $\nu$. 
However, the model cannot reproduce the observed position of the peak of the synchroton emission.
IR-optical synchrotron thermal emission does not match observational data. The
Comptonization bump is too weak, due to a weak input emission. X-ray
emission is well fitted to the quasi-stationary spectrum, but shows
no variability when the accretion rate changes in the inner flow.

Spectral indexes allow for more qualitatively comparison of the theoretical 
model and the observational data. 
Table~\ref{tab:1} lists the calculated
spectral slope, in different energy bands.
For purely thermal particles, our calculations show that the low energy radio spectral
index changes when the accretion rate changes.  
Generally, the higher the accretion rate, the steeper the spectrum is at
$\nu$ lower than the synchrotron peak. This is due to the fact
that for higher accretion rates, the synchrotron emissivity
increases. For example in the C state, there is no inner cusp, 
and most of the low energy radio photons escape
from the torus. In the B state, this energy region is dominated by the
photons coming out from the accreting cusp. The emissivity in this
range grows and $\alpha_{sp}$ steepens.  In comparison to the
observations, the  value of $\alpha_{sp}$ in the pure thermal model, is still a bit too high,
nevertheless it is much lower than in the semi-spherical models like ADAF or
spherical accretion (in ADAF, $\alpha_{sp} \sim 1.3 $, Narayan et al. 1998).

The modeled X-ray photon index in the energy band 2-10 keV for thermal
emission is constant for B, C and D states. Its value is too low to
fit the measured power law photon index in quiescence emission. 
In our thermal model this part of the spectrum is due to bremsstrahlung
emission (or power law with $\Gamma$=1.1), 
which does not agree with the interpretation of observational
data that the bremsstrahlung emission is produced 
by the 2 keV plasma (or power law with $\Gamma$=2.5).
Our model could agree with the thermal bremsstrahlung from the 2 keV plasma,
if we were to assume a much lower mass accretion rate and used
a large computational domain. We note 
that the outer radius of our
grid is $10^3 R_{S}$ as in PB03, which is 100 times
smaller than a field of view of spectroscopic observations of Sgr A*
($1.5 ''=2.4 \times 10^{5} R_{S}$).  To make our model more
self-consistent, we would have to perform our calculations over
a large range of radii, and then
compare them with the observations. This would increase
X-ray emissivity because the integration would be dominated by the
outer parts of the flow. Thus, to compensate for this increase in emissivity,
we would have to decrease $\dot{M}_a$, which would affect the synchrotron 
part of the radiation spectrum. To reduce the discrepancy, one
would also have to compute MHD models in fully 3D. Non axi-asymmetric
effects likely contribute to the time-variability of Sgr~A* and also
to the overall strength and topology of the magnetic field. We plan on 
carry out 3D MHD in near future.

In the case of a nonthermal emission, the general behavior of the radio
spectral index is similar to the thermal radio
emission. The difference is that nonthermal electrons in general produce 
a flatter spectral index in comparison to the thermal model. This
is due to the fact that a photon distribution function is not as
centered as in the thermal case. Photons in partially nonthermal
plasma are created at slightly larger distances than in thermal
plasma.  The best agreement with the data is obtained for the accretion
state C.

Our nonthermal models in X-ray energy band, agree with the data for a
weaker flare with $\Gamma=1.3$ that can be explained by models with
$\gamma_{\rm{max}}=10^6$, and the strongest flare with $\Gamma=2.5$ that can
be explained well by models with $\gamma_{\rm{max}}=10^5$.  These results
would suggest that $\gamma_{\rm{max}}$ changes are not necessarily
correlated with the flare strength.  The flare is not
extended (Quataert 2003), contrary to the quiescent emission, which also supports
our suggestion of flares being created by photons emitted by
nonthermal particles.

It seems that apart from the changes of accretion state (which can be
responsible for the IR variability and some kind of periodicity in
this frequency range), changes in $\gamma_{\rm{max}}$ are needed, to
describe the X-ray flaring behavior. To fit the observation
well, a range of $\gamma_{\rm{max}}$ should be of the order of $10^4$-$10^6$.

Summarizing, the model with a contribution of nonthermal electrons
offers a much better representation of the spectral variability of Sgr
A*, although it cannot reconstruct a few
observational points at the synchrotron peak. In addition, we would like to stress 
that although low energy radio photons are created
in the outer parts of a torus, our results are consistent with recent
measurements of Sgr A* size, which depends on $\nu$ (Shen 2006). 
In particular for $\nu > 8.5  \times 10^{12}$~Hz), 
the intrinsic size of Sgr A* is 1 AU ($1.5 \times 10^{13}$)
cm. For $\nu < 4.2 \times 10^{12}$ Hz, its size
increases by 25$\%$. Our calculations of the size of Sgr A* are $1.33
\times 10^{13}$ cm and $1.6 \times 10^{13} $cm for 
$\nu=8.5 \times 10^{12}$ $\&$ $4.2 \times 10^{12}$ Hz, 
respectively.

\subsubsection{Faraday rotation measure}

Our models can also be constrained by RM observations. 
For example, Marrone et al. (2006) found that over a period of two months,  
RM varied
within a range between $1.1 \times 10^{-5}$, and 
$23.1 \times 10^{-5} \rm{rad/m^2}$.

\begin{figure*}

\begin{center}

\includegraphics[width=0.8\textwidth,angle=90]{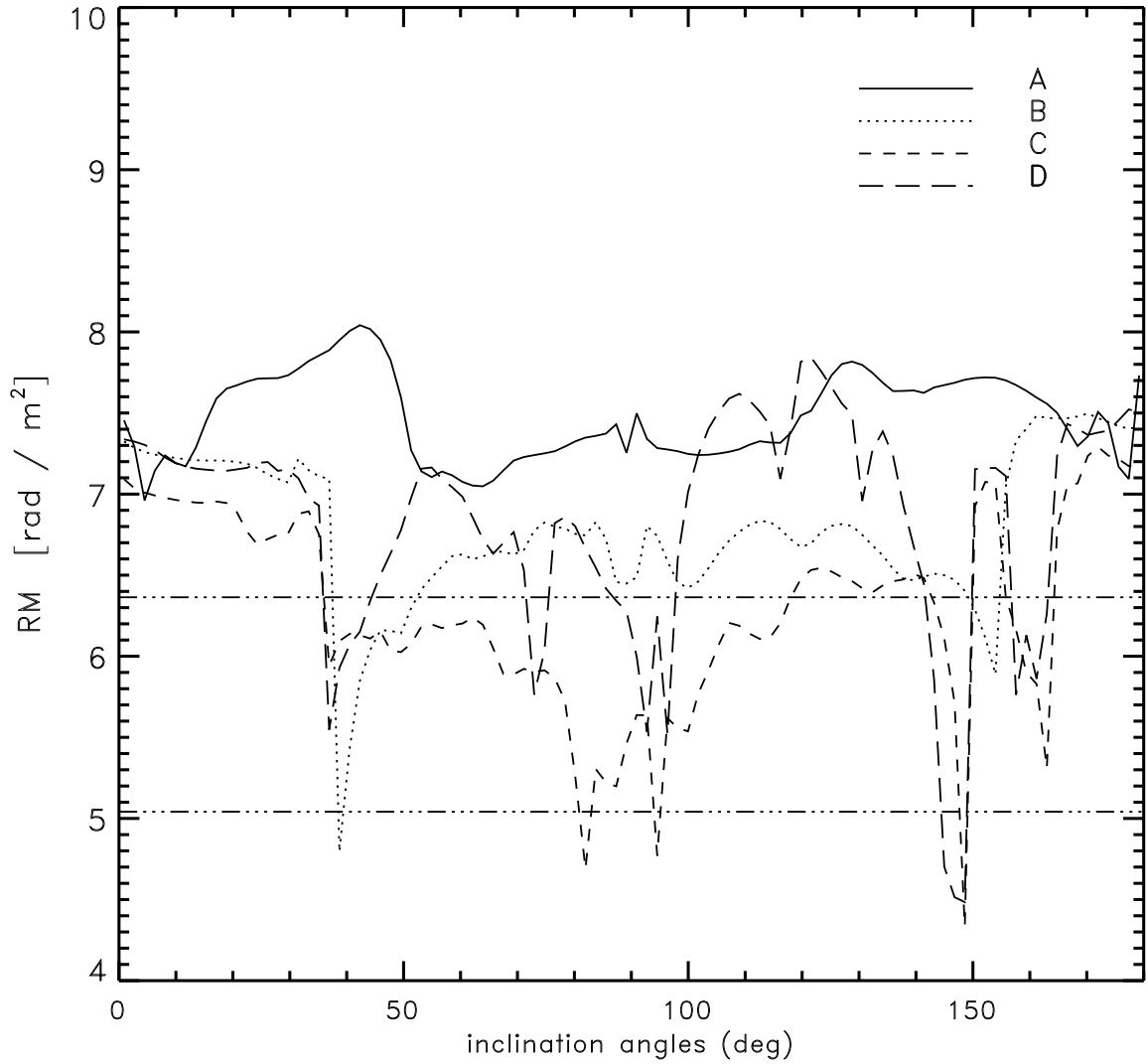}

\caption{Faraday rotation RM for four accretion states.
 x axis shows the angle of the observer, and y axis shows 
the RM values as a function of inclination predicted by models. 
Different lines show modeled RM for different time of
 simulation (A,B,C,D).
 A state is marked as solid line, B as long dash line, C as short dash line and D as dotted line.
 Two straight lines show the lower and upper limits
 from Marrone et al. (2006).}

\label{fig:7}

\end{center}

\end{figure*}

Fig.~\ref{fig:7} shows our computed RM as a function of the inclination angle,
and lower and upper limits from observations taken by Marrone et al.(2006).
The figure shows that RM depend strongly on an orientation angle of a torus
with respect to the observer. In state A, RM is very high compared to in
the state B (an accreting torus), and the measurements at various
inclinations can differ by 2 orders of magnitude.
Generally, the RM increasing with decreasing inclination angle.
This is simply because the magnetic field is highly ordered close to the
poles where there is an outflow whereas close to the equatorial plane the
magnetic field can be strongly tangled. The
model does not overpredict the observational limits either at the
equatorial plane, or around the inclinations of 30 -- 50$^\circ$,
where the inflow mixes with outflow and the region is particularly
turbulent.

\section{Conclusions}
In this paper calculations of broad band spectra predicted by MHD
simulations of low-$l$ accretion flows onto a SMBH are presented.
We studied time dependent radiative properties of MHD
flow by investigating light curves, spectra, RM and
effects of inclination.
We also studied effects of a possible non-thermal
contribution to the thermalized electrons.
We discussed  our results in the context the
observation of Sgr A*.
The summary of our main results are presented in the following:

\begin{itemize}

\item Our Monte Carlo code is quite general and allows to compute
radiative properties of relatively complex MHD flows
and include many radiative processes. 

\item 
The synthetic radiation spectrum responds to changes in 
the accretion flow in a complex non-linear way.
There is no one-to-one correspondence
between the accretion state and the predicted spectrum. For example, we
find that two states with different properties -- such as the geometry
and accretion rate -- could have relatively similar spectra.
However, we also find two very different states of accretion with very different spectra.

\item Non-thermal radiation may be required
to explain X-rays flaring observations of Galactic Center 
because thermal bremsstrahlung, that dominates 
X-ray emission, is produced at relatively large radii where
the flow changes are small and slow. 

\item Future simulations of accretion flows in LL AGN  should 
use a very large computational domain covering at least
five orders of magnitude in radius. Such simulations should capture
the dynamics of stellar winds which fuel the central BH.
This is especially important
to model properly thermal emission in X-rays but also emission 
in other wavelengths. 
In addition, future work should consider
fully 3D effects that can change the flow overall energetics and
time variability. 

\end{itemize}

\begin{acknowledgements}
We would like to thank the referee for his/her useful comments.
Thanks to Ryuichi Kurosawa and Agnieszka 
Siemiginowska for reading the manuscript.
This work was supported in part by grants 1P03D~008~29 of the Polish
State Committee for Scientific Research (KBN). 
DP acknowledges support from NASA under grant HST-AR-10305
from the Space Telescope Science Institute,
which is operated by the Association of Universities for Research
in Astronomy, Inc., under NASA contract NAS5-26555.
Partial support for this work was provided by the National Aeronautics and
Space Administration through the  Chandra award TM7-8008X issued by the
Chandra X-Ray Observatory Center, which is operated by the Smithsonian
Astrophysical Observatory for and on behalf of NASA under contract
NAS8-39073.

\end{acknowledgements}

\newpage

\newpage

\newpage

\newpage

\newpage

\newpage

\newpage

\newpage

\begin{table*}
\begin{center}

\begin{tabular}{cccccccc}
\hline 
electron distribution & energy range & $\gamma_{max}$ & State A & State B & State C & State D\\
\hline
thermal    & 1-43 GHz(a)&-    & 0.92  & 0.58 & 0.45& 0.84 \\
nonthermal & 1-43 GHz&  $10^5,10^6$ & -     &  0.33    & 0.25    &  0.61\\
thermal    & 2-10 keV(b)&-    & 1.18  & 1.17 & 1.17& 1.17\\
nonthermal & 2-10 keV&$10^5$& - & 2.1 & 2.46& 2.07  \\
nonthermal & 2-10 keV&$10^6$& - & 1.71&1.73&1.7  \\
\hline
\end{tabular}
\caption{Modeled spectral and photons indeces for different energy bands, 
accretion states, and $i=90^{\circ}$.
(a) For radio energy band we calculate spectral index $\alpha_{sp}$, 
which is defined as: $\nu F_{\nu} \sim \nu^{\alpha_{sp}+1}$. (b) 
For X-ray emission, we calculate photon index $\Gamma$,
 which is defined as: $\nu F_{\nu} \sim \nu^{2-\Gamma}$. 
We did not calculate nonthermal models for A accretion state. }
\label{tab:1}
\end{center}
\end{table*}

\newpage

\begin{table}
\begin{tabular}{cccc}
\hline 
& modeled       &        rms       & observed rms\\
\hline
& $\gamma_{max}=10^5$& $\gamma_{max}=10^6$& \\
\hline
NIR & 0.69& 0.7 &0.56\\
X-ray &0.69 & 0.68 & 0.98\\

\hline
\end{tabular}
\caption{ The modeled and observed values of the rms function.}
\label{tab:2}
\end{table}

\end{document}